\documentclass[journal,transmag]{IEEEtran}
\ifCLASSINFOpdf
  \usepackage[pdftex]{graphicx}
\else
  \usepackage[dvips]{graphicx}
\fi
\usepackage{amsmath}
\usepackage{algorithm}
\usepackage{algorithmic}

\usepackage{array}

\ifCLASSOPTIONcompsoc
 \usepackage[caption=false,font=normalsize,labelfont=sf,textfont=sf]{subfig}
\else
 \usepackage[caption=false,font=footnotesize]{subfig}
\fi

\usepackage{stfloats}

\usepackage{bm}
\usepackage{listings}
\usepackage{url}
\usepackage{amsfonts,amssymb}

\begin{document}
%
\title{Blind Inverse Gamma Correction with Maximized Differential Entropy}



\author{\IEEEauthorblockN{Yong Lee\IEEEauthorrefmark{1},	
Shaohua Zhang\IEEEauthorrefmark{2}, 
Miao Li\IEEEauthorrefmark{2}, and
Xiaoyu He\IEEEauthorrefmark{3*}}

\IEEEauthorblockA{\IEEEauthorrefmark{1} Department of Computer Science, School of Computing, National University of Singapore, Singapore}

\IEEEauthorblockA{\IEEEauthorrefmark{2}Wuhan Cobot Technology Co., Ltd., Wuhan, People's Republic of China}
\IEEEauthorblockA{\IEEEauthorrefmark{3}Department of Endocrinology, Tongji Hospital, Tongji Medical College,\\ Huazhong University of Science and Technology, Wuhan, People's Republic of China}
\thanks{Manuscript received Jan X, 2020; revised Jan X, 2020. Corresponding author Xiaoyu He(399263800@qq.com).}}

%



\IEEEtitleabstractindextext{%
\begin{abstract}
Unwanted nonlinear gamma distortion frequently occurs in a great diversity of images during the procedures of image acquisition, processing, and/or display. And the gamma distortion often varies with capture setup change and luminance variation. Blind inverse gamma correction, which automatically determines a proper restoration gamma value from  a given image, is of paramount importance to attenuate the distortion. For blind inverse gamma correction, an adaptive gamma transformation method (AGT-ME) is proposed directly from a maximized differential entropy model. And the corresponding optimization has a mathematical concise closed-form solution, resulting in efficient implementation and accurate gamma restoration of AGT-ME. Considering the human eye has a non-linear perception sensitivity, a modified version AGT-ME-VISUAL is also proposed to achieve better visual performance. Tested on variable datasets, AGT-ME could obtain an accurate estimation of a large range of gamma distortion (0.1 to 3.0), outperforming the state-of-the-art methods. Besides, the proposed AGT-ME and AGT-ME-VISUAL were applied to three typical applications, including automatic gamma adjustment, natural/medical image contrast enhancement, and fringe projection profilometry image restoration. Furthermore, the AGT-ME/ AGT-ME-VISUAL is general and can be seamlessly extended to the masked image, multi-channel (color or spectrum) image or multi-frame video, and free of the arbitrary tuning parameter. Besides, the corresponding Python code (https://github.com/yongleex/AGT-ME) is also provided for interested users.

\end{abstract}

\begin{IEEEkeywords}
Blind Inverse Gamma Correction, Differential Entropy, Gamma Estimation
\end{IEEEkeywords}}

\maketitle

\IEEEdisplaynontitleabstractindextext

%
\IEEEpeerreviewmaketitle

\section{Introduction}
\label{intro}
\IEEEPARstart{E}{}stimating the amount of gamma distortion is of great importance for contrast enhancement \cite{rahman2016adaptive}, image preprocessing\cite{babakhani2015automatic}, image-based measuring profilometry\cite{li2011gamma}, "one shot profilometry" gamma correction\cite{garcia2019simultaneous}, automatic gamma adjustment during image capture\cite{toshinobu1992auto}, etc.
Image blind inverse gamma correction targets to automatically estimate the amount of gamma distortion, in the absence of environmental luminance information or device capturing settings\cite{farid2001blind, babakhani2015automatic}. That is to say, we cannot compute an appropriate gamma correction value from the analysis of the imaging systems, including luminance, lens configuration, properties of an imaging device.  

Specifically, image blind inverse gamma correction can be regarded as a challenging inverse problem in mathematics, because both the ideal distortion-free image and distortion gamma value need to be determined solely from a distorted picture. Thus, choosing a proper prior is paramount for solving this inverse problem, and different priors lead to various methods \cite{farid2001blind,lin2004radiometric,cao2010forensic,babakhani2015automatic}, detailed in Section \ref{related_works}. The challenges also come from the engineering aspects.  Regarding on-line gamma adjustment or image preprocessing applications, the algorithm should have low computational complexity/ energy consumption\cite{babakhani2015automatic} and be easily implemented with the simple embedded circuit of an imaging device\cite{toshinobu1992auto}.  

Shannon information entropy is a prestigious metric to measure the total amount of information content present in an image, and has been used for various aspects of image processing, including noise removal \cite{chang2010entropy}, deblurring \cite{carasso1999linear},  speckle removal \cite{rajalaxmi2014entropy}, contrast enhancement \cite{celik2014spatial}, and phase recalibration \cite{gull1984maximum}.  Maximum entropy has been proved to be an effective image prior because most good images should contain sufficient information. Related literature reported that a better image (well-focused, blur-free, good-light) actually has relatively higher entropy, and see also \cite{smith2013maximum} for general discussions. As far as we know, the maximum entropy prior has not been used to the gamma distortion estimation or blind inverse gamma correction problem. 

\begin{figure*}[!t]
\centering
\subfloat[]{\includegraphics[width=6.4in]{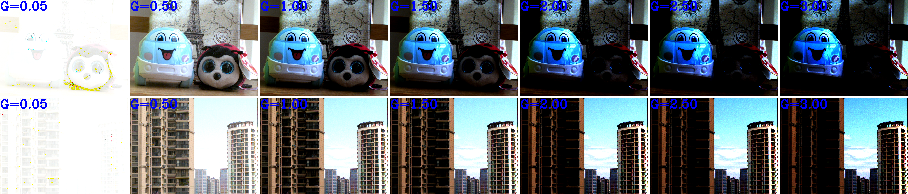}}
\hfil
\subfloat[]{\includegraphics[width=3.2in]{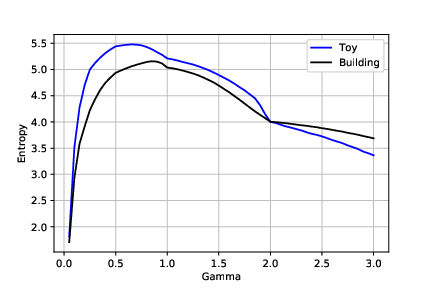}}
\subfloat[]{\includegraphics[width=3.2in]{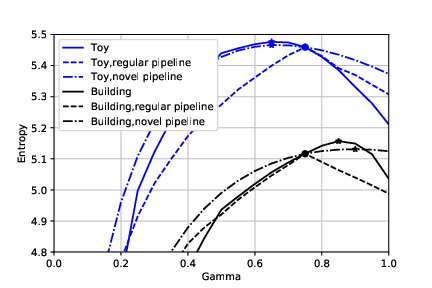}}
\caption{ Toy images(upper) and building images (bottom) captured by an industrial camera(HKvision, MV-CE003-20GC) with varing gamma (denoted as $G$)(a), and the corresponding entropy values (b). Given the image at gamma $0.75$ (circle marker), two pipelines predict the entropy value with different gamma settings, the maximal entropy values are marked with star markers(c).}
\label{fig_1}
\end{figure*}

There is a \em{quantized entropy decrease barrier}\em{} that prevents the direct utilization of maximum entropy prior to the blind inverse gamma correction. A regular pipeline to computing the entropy of a gamma transformed image is: original image $\stackrel{power-law}\longrightarrow$ float image $\stackrel{quantize}\longrightarrow$ transformed image $\rightarrow$ entropy. Due to the peak-gap pattern\cite{cao2010forensic} from quantization step, the entropy of a transformed image always deceases, as in Fig.\ref{fig_1} (c). We call this phenomina as \em{quantized entropy decrease barrier}\em{}.


Despite the \em{quantized entropy decrease barrier}\em{}, we argue that the classical and effective maximum entropy prior should be worked well for blind inverse gamma correction. Therefore, we first verified that gamma distortion will lead to entropy decrease, i.e., a gamma-distortion free image has the maximum entropy, see Fig.\ref{fig_1} (a, b). Based on this observation, we thus proposed an adaptive gamma transformation algorithm with maximized entropy, named as AGT-ME, to estimating the best amount of gamma correction automatically. Considering the human eye has a non-linear response to brightness, a modified algorithm, called as AGT-ME-VISUAL, is also proposed for more attractive visual performance. The main contributions of this work:
\begin{enumerate}
\item Despite the \em{quantized entropy decrease barrier}\em{}, this manuscript successfully formulates the blind inverse gamma correction problem with maximum entropy prior. This maximum entropy assumption provides a solid guarantee of accuracy in theory.
\item Based on differential entropy and change-of-variables formula, a concise closed-form solution to the problem is obtained, resulting in an exact (non-approximate) algorithm. Due to the closed-form solution, we provide efficient algorithm implementations and could further analyze the algorithm characteristics.
%
\item Comprehensive experimental results provided quantitative and qualitative assessments, thus illustrated the AGT-ME and AGT-ME-VISUAL's effectiveness, efficiency. And we also tested the performance on automatic gamma adjustment, image contrast enhancement, and fringe projection profilometry image restoration, which demonstrated a wide application range.
\end{enumerate}

The rest of the paper is structured as follows. Some related works are given in Section \ref{related_works}. Section \ref{methodology} details the proposed AGT-ME and AGT-ME-VISUAL methods from problem formulation to algorithm implementation. Section \ref{experiments} tests our algorithm on both synthetic datasets and several image datasets with comparisons to classical algorithms, followed by three relevant applications. A few concluding comments are drawn in Section \ref{conclusion}.

\section{Related works}
\label{related_works}

Blind inverse gamma correction is firstly introduced by Farid \cite{farid2001blind}. He noticed that gamma distortion introduces higher-order correlations (high-order harmonics) in the frequency domain, which can be measured with a bicoherence metric in polyspectral analysis. The restoration gamma value is located at where the bicoherence metric is minimized. Thus, Farid adopted the minimum bicoherence prior to this problem. This minimum bicoherence prior gains the insight into the gamma correction for later studies\cite{li2011gamma}. However, this prior is computational heavy, the approximations (scan lines)  and numerical optimization (search strategy) degrade the performance.



Lin et al. \cite{lin2004radiometric} found that a non-linear gamma distortion warps the image linear irradiance colors into a non-linear distribution in the edge regions, and utilized a total distance to measure the non-linearity. After selecting appropriate edge information, a Bayesian approach is used to perform the gamma calibration. This prior \cite{lin2004radiometric} is not a general assumption, the pixels around an edge do not necessarily satisfy the linear relationship. And, extra work to find the edge regions is needed.

Cao et al. \cite{cao2010forensic} observed that the histogram of a gamma distorted image has peak and gap bins pairs, i.e., peak-gap pattern or peak-gap fingerprint. This peak-gap fingerprint distribution prior is uniquely determined by gamma distortion amount and independent of the image's content. However, the raw images (captured by a camera) don't have peak-gap fingerprints even if an unreasonable gamma is set, and the image peak-gap fingerprint prior is also a volatile feature, which is missing after further processing (blur, for instance). 

Vazquez-Corral and Bertalm{\'\i}o \cite{vazquez2015simultaneous} exploited the structure of the standard in-camera processing pipeline, and estimate the gamma distortion by employing multiple aligned images from the same scene, i.e., the same position pixels from two or more distortion-free images of the same scene should have a linear relationship prior. The minimization is performed via a time-consuming brute-force search. Note that this method\cite{vazquez2015simultaneous} require multi-view color images. And therefore this multi-view prior is not suitable for a single image gamma correction.

Babakhani and Zarei \cite{babakhani2015automatic} supposed a gamma that changes the average of brightness to $\frac{1}{2}$, then the correction gamma $\gamma$ is calculated from $A^\gamma=\frac{1}{2}$, where $A$ is the average of the brightness. This "$\frac{1}{2}$ - average" prior is intuitive and can be easily implemented as an efficient algorithm. However, "$\frac{1}{2}$ - average" prior has an average shift problem, i.e., the average of a distorted image $A_d$ is not equal to the gamma power of the average of the original image $A_o$, $A_d\neq A_o^\gamma$. In addition, there is no guarantee that a distortion-free image has average brightness $0.5$. Note that $A$ is the first-order statistics of image PDF, only represents part of whole image information. Thus, this "$\frac{1}{2}$ - average" prior is not an exact assumption.

Mahamdioua and Benmohammed \cite{mahamdioua2017new} defined a feature vector of means/variances of image rows and columns, which is expected to tell the distortion-free and distorted image apart. Specifically, a series of target vectors from good-light images is firstly constructed, and then a range of inverse gamma values are applied to the input image, find the best gamma value with which the feature vector of transformed image has the minimal distance to the target vectors. The corresponding model is slowly optimized by trial and error. This is an interesting method that tries to learn a prior from distortion-free images, but the learned prior is complex and difficult to use as well.


In the area of fringe-projection profilometry (FPP), several customized adaptive gamma correction methods \cite{guo2004gamma,liu2010gamma,hoang2010generic,li2011gamma,garcia2019simultaneous} were proposed to address gamma distortion problem with the spatial sinusoidal prior of fringe images. It is complicated to address the fringe phase extraction as well as gamma correction simultaneously. Anyway, an adaptive gamma correction (similar to camera automatic exposure\cite{zhang2006automatic}), which needs a fast and accurate gamma estimation algorithm, should improve the measurement accuracy at the initial image acquisition phase. 

Table.\ref{table_2} compares thoese methods in terms of theoretical basis, solution property, and other key characteristics. From the table, one can find that the proposed methods have solid theoretical basis and nice properties. Hence, AGT-ME and AGT-ME-VISUAL are more likely to give better performance (which we will show in Section \ref{experiments}).

\begin{table*}[!t]
\renewcommand{\arraystretch}{1.3}
\caption{Comparison  of different methods for blind inverse gamma correction.}
\label{table_2}
\centering
\begin{tabular}{|c|l|c|c|c|c|c|}
\hline
 \bf{Method}                                  & \bf{Theoretical basis} & \bf{Closed solution} & \bf{Speed} & \bf{Single gray image} & \bf{Information source} & \bf{Comments}\\
\hline
BIGC\cite{farid2001blind}                     & bicoherence          & $\times$        & slow  & \checkmark        & image part (scan lines)& -       \\
\hline
Lin\cite{lin2004radiometric}                  & color distribution   & $\times$        &  -    & $\times$          & image part (edges)&require color image \\
\hline
Cao\cite{cao2010forensic}                     & histogram fingerprint& $\times$        &  -    & \checkmark        & part of histogram &peak-gap is volatile \\
\hline
Vazquez-Corral\cite{vazquez2015simultaneous}  & multiple views       & $\times$        & slow  & $\times$          & image pair&require image alignment  \\
\hline
FPPs\cite{guo2004gamma,li2011gamma,liu2010gamma} &special sinusoidal prior&$\times$    & -     & $\times$          & - &special application\\
\hline
CAB\cite{babakhani2015automatic}              & average of brightness& \checkmark      & fast  & \checkmark        & whole image&arbitrary assumption \\
\hline
AGT-ME (ours)                                 & maximum entropy      &\checkmark       & fast  & \checkmark        & whole image&accurate, fast\\
\hline
AGT-ME-VISUAL(ours)                           & maximum entropy      &\checkmark       & fast  & \checkmark        & whole image&fast, designed for human  \\
\hline
\end{tabular}
\end{table*}




\section{METHODOLOGY}
\label{methodology}

\subsection{Concepts Introduction}
\em{Gamma transformation}\em{}.
The nonlinear gamma distortion and correction are always modeled as a pixel-wise power-law function with a single parameter gamma $\gamma$, i.e, gamma transformation in digital image processing \cite{gonzalez2002digital}.  The following Eq.(\ref{eq_1}) recalls the gamma transformation $G(\cdot)$ of an image $\bm{I}$. 
\begin{equation}
\label{eq_1}
g_m=f(u_m)= u_m^{\gamma}
\end{equation}
where a gray-scale image $\bm{I}$  consists of $M$(width $\times$ height) individual pixels, the subscript $m\in \{0,1,...,M-1\}$ and $u_m\in [0,1]$ respectively refer to the pixel index and intensity, and $g_m$ denotes the image intensity at the same position after gamma transformation. The transformation function can be controlled by varying the gamma $\gamma$ values, and thus it can deal with bright and dimmed images by selecting a proper $\gamma$. 

\em{Differential entropy}\em{}\cite{cover2012elements} is a generalised continuous variant of the discrete Shannnon entropy, with definition below:
\begin{equation}
H(\bm{I}) = -\int_0^1p_{\bm{I}}(u)\log_2[p_{\bm{I}}(u)]du
\label{eq_2}
\end{equation}
where $p_{\bm{I}}(u)$ is the PDF of image $\bm{I}$, which is approximated by the division of discrete histogram and pixel number $M$. The difference between Shannon entropy and differential entropy will not be discussed here. In this special case, image intensity $u$ is normalized to range $(0,1)$, differential entropy is equivalent to the limit of Shannon entropy, and used as a metric to assess image gamma distortion.


\em{Change of variable rule}\em{}. After gamma transformation (Eq.(\ref{eq_1})), the density functions of transformed image $G(\bm{I})$ can be deduced from PDF of image $\bm{I}$ in a continuous view with the help of probability analysis \cite{durrett2019probability}(Exercises 1.2.5), i,e., change-of-variables formula.
\begin{equation}
\begin{split}
p_{\bm{G}}(g) &= p_{\bm{I}}(f^{-1}(g))|\frac{\partial}{\partial g}f^{-1}(g)|\\
 & = p_{\bm{I}}(f^{-1}(g))\frac{1}{\gamma}(f^{-1}(g))^{1-\gamma}
\end{split}
\label{eq_3}
\end{equation}
where $p_{\bm{G}}(g)$ is the probability density function of image $G(\bm{I})$. This formula means that we can directly predict the PDF (histogram) of the transformed image without explicitly applying gamma transformation to the original image.

\subsection{Formulation, Solution and Characteristics}
Base on the  maximum entropy prior (Sect. \ref{intro}), the best restoration gamma $\gamma^*$ should lead to a maximum entropy of transformed image, which can be formulated as: 
\begin{equation}	
\gamma^*=\arg \max_{\gamma} H(G(\bm{I}))
\label{eq_4}
\end{equation}
Here, we introduce an auxiliar negative entropy loss $J(\gamma)=- H(G(\bm{I}))$, Eq.(\ref{eq_4}) is thus transformed to 
\begin{equation}	
\gamma^*= \arg \min_{\gamma} J(\gamma)
\label{eq_5}
\end{equation}
with 
\begin{equation}	
J(\gamma) =  \int_0^1p_{\bm{G}}(g)\log_2[p_{\bm{G}}(g)]dg
\label{eq_6}
\end{equation}
Take the Eq.(\ref{eq_1}) and Eq.(\ref{eq_3}) in Eq.(\ref{eq_6}), we can get:
\begin{equation}
\begin{split}
J(\gamma)& = \int_0^1 	p_{\bm{I}}(u)\log_2[p_{\bm{I}}(u)\frac{1}{\gamma}u^{1-\gamma}]du
\end{split}	
\label{eq_7}
\end{equation}
Note that Eq.(\ref{eq_7}) is a novel pipeline (image  $\rightarrow$ PDF $\rightarrow$ transformation $\rightarrow$ entropy) that predicts the negtive entropy of a transformed image without computing $G(\bm{I})$, as shown in Fig.\ref{fig_1}(c). Due to  $\frac{\partial^2 J(\gamma)}{\partial \gamma^2}>0$,  let $\frac{\partial J(\gamma)}{\partial \gamma}=0$, a concise closed-form solution of the problem (Eq.(\ref{eq_4})) is obtained.
\begin{equation}
\gamma^* = -\frac{1}{\int_0^1p_{\bm{I}}(u)\ln(u)du}
\label{eq_8}
\end{equation}
where $\ln(\cdot)$ means the natural logarithm. Note that other metrics, such as $-\int_0^1p^2(t)dt$,  don't have such a concise closed-form solution.
Observing the solution in Eq.(\ref{eq_8}), five interesting characteristics can be precisely deduced.
\begin{enumerate}
	\item $0<\gamma^*<\infty$.
	\item if $p_{\bm{I}}(u)=1$ is a uniform distribution, the $\gamma^*=1$.
	\item if $u_m=c \in (0,1)$, the $g_m=c^{\frac{-1}{\ln(c)}}=e^{-1}$.
	\item entropy gain $H(G(\bm{I})) -H(\bm{I})= \log_2(\gamma^*)+\frac{1}{\gamma^*}-1\geq 0$.
  \item a certain similarity  with result $-\frac{\ln(2)}{\ln [\int_0^1p_{\bm{I}}(u)udu]}$ in ref\cite{babakhani2015automatic}.
\end{enumerate}

\subsection{Implementation}
\subsubsection{Computational Details}
The closed-form solution (Eq.(\ref{eq_8})) still needs to compute the PDF $p_{\bm{I}}(u)$ of input image $\bm{I}$. We found that it is feasible to perform discrete computing in pixel-domain for the expectation $\mathop{\mathbb{E}}_{u\sim p_{\bm{I}}(u)} \ln(u)$. 
\begin{equation}
\gamma^* = -\frac{1}{\frac{1}{M}\sum_{m=0}^{M-1} \ln(u_m)}
\label{eq_9}
\end{equation}
and recall $u_m\in [0,1]$ the pixel intensity of original image. The closed interval $[0,1]$ suffers boundary calculation problem ($\ln(0)$ or $\frac{1}{0}$ ). Thus, a trick is to normalize the image intensity value to an open interval $(0,1)$. For instance, a 8-bit gray image with 256 intensity levels could be normalized with Eq.(\ref{eq_10}) in this work.

\begin{equation}
u_m = \frac{l_m+0.5}{256}
\label{eq_10}
\end{equation}
where $l_m\in\{0,1,2,...,255\}$ is the  intensity level of a  8-bit gray-scale image.

\subsubsection{Color Image}
To deal with a color image, one of the following tricks is feasible. 
\begin{itemize}
	\item extend the computation of Eq.(\ref{eq_9}) directly to all channels of a color image;
	\item compute restoration $\gamma^*$ from corresponding gray-scale image, and apply gamma transformation on each channel of RGB color space with a common estimation $\gamma^*$;
	\item treat each channel of RGB image as a gray-scale image, and perform adaptive gamma correction independently;
	\item conduct adaptive gamma correction in a specific color space channel.  Because the luminance value (V) and color information (Hue and Saturation) are decoupled in HSV color space, the gamma correction can be done on V-channel of HSV color space \cite{tsai2013adaptive}. This strategy is selected in this work since it will not cause extra color distortion during gamma correction.
\end{itemize}

\subsubsection{Region of Interesting}
Mask operation is also supported in AGT-ME method. The optimal gamma value can be computed only in the region of interesting.
\begin{equation}
\gamma^* = -\frac{1}{\frac{1}{N}\sum_{m\in \Omega} \ln(u_m)}
\label{eq_11}
\end{equation}
where $\Omega$ denotes the mask area which is always recorded in a mask image $\bm{M}$,  with valid pixel number $N$. 

\subsubsection{Better Visualization}
The corrected image with optimal gamma value $\gamma^*$ tends to have a uniform PDF, similar to histogram equalization. Because the human eye has a nonlinear response to brightness, a power-law encoding with a gamma value of $2.2$ produces a much more even distribution of quantization steps \cite{reinhard2010high}. To achieve a better visual performance, a visual gamma $\gamma^*_v$ could be employed.
\begin{equation}
\gamma^*_v =\gamma^*/2.2
\label{eq_12}
\end{equation}

We find that every valid pixel is contributed to the gamma estimation (Eq.(\ref{eq_9})) instead of a small fracture of sampled pixels (scan-lines\cite{farid2001blind} or pixels in edge regions\cite{lin2004radiometric}). And the blind gamma estimation is very efficient by computing Eq.(\ref{eq_9}), which could be easily extended to masked, multi-channel color image and multi-frame video. To this end, we can summary our AGT-ME as Algorithm \ref{algorithm_1}. Also, human visual perception sensitivity is considered by incorporating a visual gamma $\gamma^*_v$ in a modified AGT-ME algorithm, acronymed AGT-ME-VISUAL, as Algorithm \ref{algorithm_2}.


\renewcommand{\algorithmicrequire}{\textbf{Input:}}
\renewcommand{\algorithmicensure}{\textbf{Output:}}

\begin{algorithm}[!ht]
\caption{Adaptive Gamma Correction method (AGT-ME)}
\label{algorithm_1}
\begin{algorithmic}[1]
\REQUIRE a gray-scale image or V-channel of color image  $\bm{I}$,\\
          a binary mask image $\bm{M}$ (default full image).
\ENSURE gamma estimation $\gamma^*$, corrected image $G(\bm{I})$
\STATE normalize the image $\bm{I}$ with Eq.(\ref{eq_10});
\STATE compute the optimal gamma $\gamma^*$ with Eq.(\ref{eq_11});
\STATE adopt $\gamma^*$ to perform gamma correction with Eq.(\ref{eq_1});
\STATE do inverse normalization, obtain final corrected $G(\bm{I})$.
\end{algorithmic}
\end{algorithm}

\begin{algorithm}[!ht]
\caption{AGT-ME-VISUAL}
\label{algorithm_2}
\begin{algorithmic}[1]
\REQUIRE a gray-scale image or V-channel of color image  $\bm{I}$,\\
          a binary mask image $\bm{M}$ (default full image).
\ENSURE gamma estimation $\gamma_v^*$, corrected image $G(\bm{I})$
\STATE normalize the image $\bm{I}$ with Eq.(\ref{eq_10});
\STATE compute the optimal gamma $\gamma^*$ with Eq.(\ref{eq_11});
\STATE obtain the visual gamma $\gamma_v^*$ with Eq.(\ref{eq_12});
\STATE adopt $\gamma_v^*$ to perform gamma correction with Eq.(\ref{eq_1});
\STATE do inverse normalization, obtain final corrected $G(\bm{I})$.
\end{algorithmic}
\end{algorithm}





\section{Experiments}
\label{experiments}
In this section, the characteristics and accuracy of AGT-ME and AGT-ME-VISUAL methods are illustrated through comparison with reference correction methods on different signals. The reference methods are classical blind inverse gamma correction method (BIGC) \cite{farid2001blind} and correction on average of brightness (CAB) algorithm\cite{babakhani2015automatic}. To our best knowledge, only BIGC, CAB, AGT-ME, and AGT-ME-VISUAL can provide a gamma correction from a single gray-scale natural image. 
Firstly, a one-dimensional synthetic signal\cite{farid2001blind} is analyzed to gain insight into the methods. Secondly, the accuracy of gamma estimation is fully investigated by estimating the extra simulated gamma distortion of natural gray-scale images. Besides, the low computational complexity is verified by a computational cost experiment. Finally, our AGT-ME and AGT-ME-VISUAL are tested with several related applications, including automatic gamma adjustment for an imaging device, natural/medical image contrast enhancement, and FPP image gamma correction.

\subsection{One-Dimensional Signal Analysis}

\begin{figure*}[!t]
\centering
\subfloat[]{\includegraphics[width=3.2in]{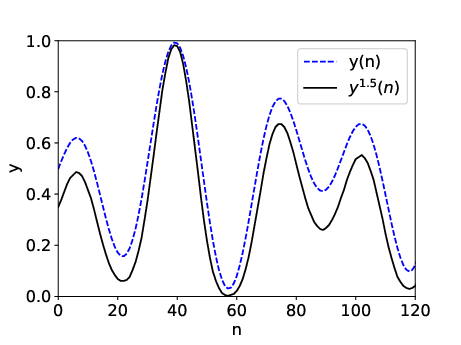}%
\label{fig_1a}}
\subfloat[]{\includegraphics[width=3.2in]{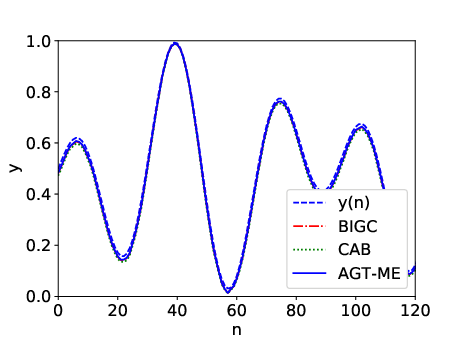}%
\label{fig_1b}}
\hfil
\subfloat[]{\includegraphics[width=3.2in]{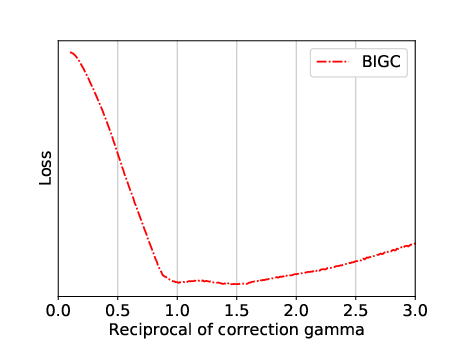}%
\label{fig_1c}}
\subfloat[]{\includegraphics[width=3.2in]{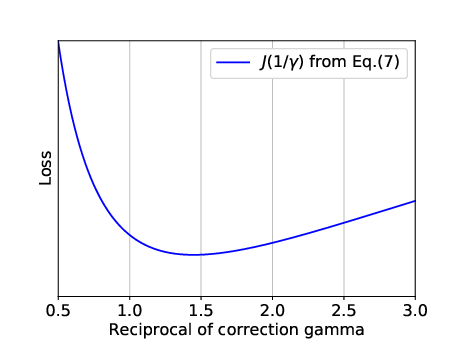}%
\label{fig_1d}}
\caption{A synthetic signal and the distorted curve(a). (b) represents the corresponding corrections. (c) and (d) are the loss curves for BIGC and AGT-ME.}
\label{fig_2}
\end{figure*}

An arbitrary 1-D signal $y(n)$ with 512 points\cite{farid2001blind} is employed here.
\begin{equation}
y(n) =  \frac{75}{255}\sin(2\pi \frac{2n}{64})-\frac{55}{255}\sin(2\pi \frac{1.3n}{64}))+\frac{127}{255}
\end{equation}
where $n \in \{0,1,...,511\}$ denotes the position. Simulated gamma distortion $(\gamma =1.5)$ are performed to $y(n)$, resulting in the distorted signal $y^{1.5}(n)$. Note that all the signals are quantized in 8-bit discrete level. Fig.\ref{fig_1}(a) depicts both $y(n)$ (blue dash line) and $y^{1.5}(n)$ (black solid line) in a 120-point length segment.
Given only the distorted $y^{1.5}(n)$ signal, BIGC and CAB  estimate the distorted gamma value as $1.4385$ and $1.3999$ respectively. Meanwhile, the AGT-ME provides a more accurate estimation $1.4478$. The slight differences are observable from the recovered signals, as shown in Fig.\ref{fig_1} (b).

Further investigation about the loss curves is also performed. A series of inverse gamma value is applied to the distorted signal $y^{1.5}(n)$,  and get the loss measure changes in Fig.\ref{fig_1}(c) and (d). The BIGC loss (bicoherence metric) and AGT-ME loss (entropy metric) have a similar overall shape, it implies that both BIGC bicoherence and AGT-ME entropy could identify the gamma distortion. Regarding this special case,  BIGC does not provide a strick convex loss curve, which means a potential risk of getting the local optimal solution. On the contrary, the negative entropy function (Eq.(\ref{eq_7})) is convex and has a unique global minimum point. 





\subsection{Gamma Estimation on Synthetic Gamma Distortion}

\begin{figure}[!t]
\centering
\includegraphics[width=3.5in]{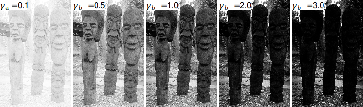}
\caption{Synthetic gamma distortions with different gamma values. $\gamma_b=1.0$ is the original image.}
\label{fig_3}
\end{figure}

\begin{figure}[!t]
\centering
\subfloat[]{\includegraphics[width=3.2in]{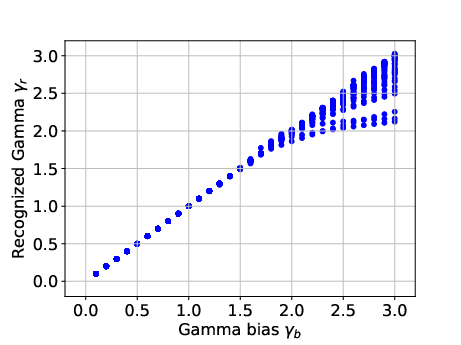}%
\label{fig_3a}}
\hfil
\subfloat[]{\includegraphics[width=3.2in]{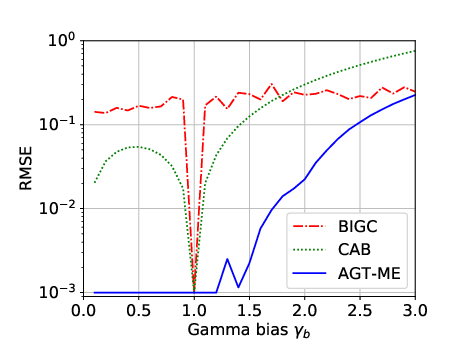}%
\label{fig_3b}}
\caption{Estimated gamma values over synthetically distorted images with AGT-ME (a). The root mean square error of the estimations (b). For a clear visualization, the RMSE value is clamped with the minimum value of $10^{-3}$.}
\label{fig_4}
\end{figure}
Natural images have more diversity than any other specific kind of image, and therefore, it is appropriate to use natural images to evaluate the accuracy of gamma estimation. Dataset BSD68 \cite{roth2005fields,martin2001database,zhang2018ffdnet}, containing 68 natural gray-scale images, is adopted due to sufficient image quantity. Unfortunately, the exact gamma (ground truth) distortion of this dataset is not available, and we have not yet found any image dataset with ground-truth gamma annotation. Fortunately, extra distortion can be unambiguously synthesized with a bias gamma value $\gamma_b$. Fig.\ref{fig_2} displays the synthetic distorted images with different $\gamma_b$. Providing the image pairs before and after the gamma transformation, the relative extra distortion thus can be recognized as $\gamma_r$ by comparing the gamma estimations of the pairs. Providing $\gamma_b$ and the recognized gamma difference $\gamma_r$, the accuracy can then be evaluated with root mean square error (RMSE).
\begin{equation}
RMSE = \sqrt{\frac{\Sigma_{k=1}^K (\gamma_{r,k}-\gamma_{b})^2}{K}}
\label{eq_14}
\end{equation}
where $k\in\{1, 2, ..., K\}$ is the image index of dataset BSD68 with size $K=68$.


Fig.\ref{fig_3} (a) scatters the $(\gamma_b, \gamma_r)$ pair with $\gamma_b$ ranges from 0.1 to 3.0 at interval 0.1, and each point reports an AGT-ME result of the $30\times68$ synthetic distorted images. Most of the scatter points located close to the straight line $(\gamma_r=\gamma_b)$, which means the $\gamma_r$ is quite accurate for the majority of the cases. We also measured the RMSE for different actual gamma bias $\gamma_b$, shown in Fig.\ref{fig_3}(b). At a special position $\gamma_b=1.0$, three methods have zero-error due to the same image before and after extra gamma distortion. Excluding this special position, the degree of error dispersion increases with the value $\gamma_b$. The RMSE result indicates that our AGT-ME performs much better than the BIGC method or CAB approach in the whole investigation range, and BIGC, CAB, and AGT-ME have an average RMSE value $0.2016$, $0.2420$ and $0.0439$ respectively. Note that, about $27.3\%$ outliers($|\gamma_r-\gamma_b|>0.5$) are excluded for BIGC method.



\subsection{Algorithm Complexity Analysis and Execution Time Cost}

\begin{table}[!t]
\renewcommand{\arraystretch}{1.3}
\caption{Average execution time ($ms$) of different methods}
\label{table_1}
\centering
\begin{tabular}{|c|c|c|c|}
\hline
 Image size  & BIGC & CAB &AGT-ME\\
\hline
$256*256$    & 259.2  &  3.4 & 5.2\\
\hline
$512*512$    & 637.0  &  14.0&  20.1\\
\hline
$1024*1024$    &  1422.1  &63.3 & 82.6\\
\hline
$2048*2048$    &  3144.5  &216.0 & 310.4\\
\hline
\end{tabular}
\end{table}

The proposed AGT-ME method has a computational complexity $\mathcal{O}(M)$(Eq.(\ref{eq_9})), resulting in a theoretically efficient blind inverse gamma correction algorithm. The BIGC, CAB, and AGT methods were tested experimentally as well using Python 3.5 on a 2.30 GHz i5-8300H laptop computer with RAM 8.00 GB. Experimental results (Table.\ref{table_1}) with varied input image size demonstrated that our algorithm is fast as expected. Note that, AGT-ME could be implemented faster via efficient C++ and/or look-up table trick \cite{guo2004gamma}. 

\subsection{Applications}
\subsubsection{Automatic Gamma Setting for Digital Camera}
 
\begin{figure*}[!t]
\centering
\subfloat[Init, \protect\\ $\gamma(0.75), H(5.460)$ ]{\includegraphics[width=1.1in]{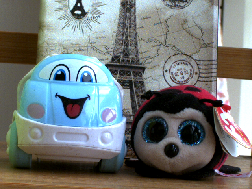}}
\vspace{0.001in}
\subfloat[Max entropy, \protect\\ $\gamma(0.65), H(5.477)$]{\includegraphics[width=1.1in]{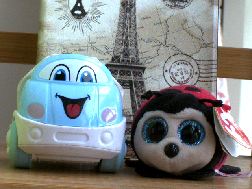}}
\vspace{0.001in}
\subfloat[BIGC, \protect\\ $\gamma(2.00), H(4.000)$]{\includegraphics[width=1.1in]{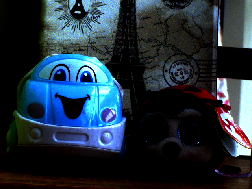}}
\vspace{0.001in}
\subfloat[CAB,  \protect\\$\gamma(0.70), H(5.474)$]{\includegraphics[width=1.1in]{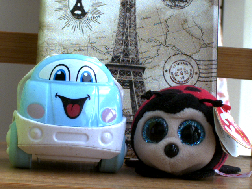}}
\vspace{0.001in}
\subfloat[AGT-ME, \protect\\ $\gamma(0.65), H(\bm{5.477})$]{\includegraphics[width=1.1in]{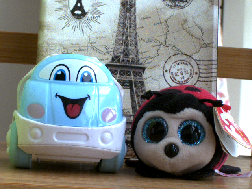}}
\vspace{0.001in}
\subfloat[AGT-ME-VISUAL, \protect\\ $\gamma(0.30), H(5.122)$]{\includegraphics[width=1.1in]{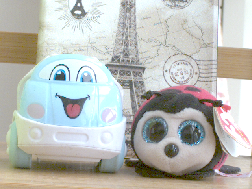}}
\hfil
\subfloat[Init, \protect\\ $\gamma(0.75), H(5.116)$ ]{\includegraphics[width=1.1in]{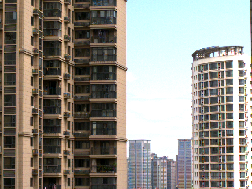}}
\vspace{0.001in}
\subfloat[Max entropy, \protect\\ $\gamma(0.85), H(5.157)$]{\includegraphics[width=1.1in]{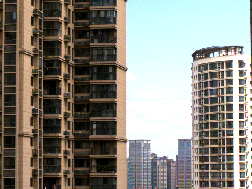}}
\vspace{0.001in}
\subfloat[BIGC, \protect\\ $\gamma(3.00), H(3.687)$]{\includegraphics[width=1.1in]{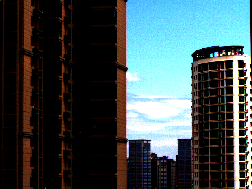}}
\vspace{0.001in}
\subfloat[CAB,  \protect\\$\gamma(0.90), H(\bm{5.149})$]{\includegraphics[width=1.1in]{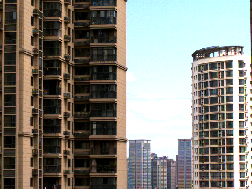}}
\vspace{0.001in}
\subfloat[AGT-ME, \protect\\ $\gamma(0.90), H(\bm{5.149})$]{\includegraphics[width=1.1in]{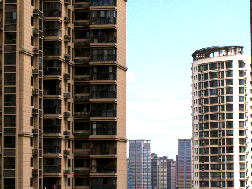}}
\vspace{0.001in}
\subfloat[AGT-ME-VISUAL, \protect\\ $\gamma(0.40), H(4.746)$]{\includegraphics[width=1.1in]{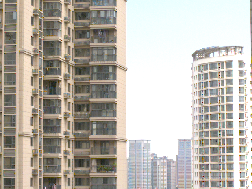}}
\caption{Automatic gamma adjust setting. (a) and (g) are the initial image. (b) and (h) are the images with maximum entropy. (c-f) and (i-l) are the results of BIGC, CAB, AGT-ME, and AGT-ME-VISUAL. The $\gamma(\cdot)$ is the camera gamma setting and the $H(\cdot)$ denotes the corresponding image entropy.}
\label{fig_5}
\end{figure*}

Digital camera sensors usually have a linear response to the input light. In the process of rendering linear raw data to conventional RGB data, gamma transformation will be performed. Automatic gamma adjustment is as important as automatic exposure to obtain a good image. The experiment is based on an image dataset captured with an industrial camera (HKvision, MV-CE003-20GC\footnote{\url{https://www.hikrobotics.com/vision/visioninfo.htm?type=42&oid=2121}}) with gamma from $0.05$ to $3.00$ with an interval of $0.05$ (Fig.\ref{fig_1}). Providing an initial image captured by a default gamma value $0.75$ (Fig.\ref{fig_5}(a) and (g)), different methods automatically adjust the gamma value with results displayed in Fig.\ref{fig_5}.

The BIGC method did not set a proper gamma in both "toy" and "building" cases, resulting in dimmed images with low entropy. The CAB and AGT-ME performed similarly in visualized results, but AGT-ME achieved a larger entropy in the "toy" case. Recall the entropy prediction depicted in the Fig.\ref{fig_1}(c), the AGT-ME method agrees well with the ground truth, which explains the performance of automatic gamma adjustment. The result images of AGT-ME-VISUAL (Fig.\ref{fig_5}(f) and (l)) look like a little over bright, with the entropy slightly decreased. However, the fine details in the dark areas (muppet eye and building windows) become more clear from human perception aspect. Totally speaking, the AGT-ME could achieve the maximum entropy goal for automatic gamma adjustment, and AGT-ME-VISUAL displays a more informative image for humans.

\subsubsection{Gamma Correction for Natural and Medical Image}

\begin{figure*}[!t]
\centering
\subfloat[Flower 1]{
\includegraphics[width=7.2in]{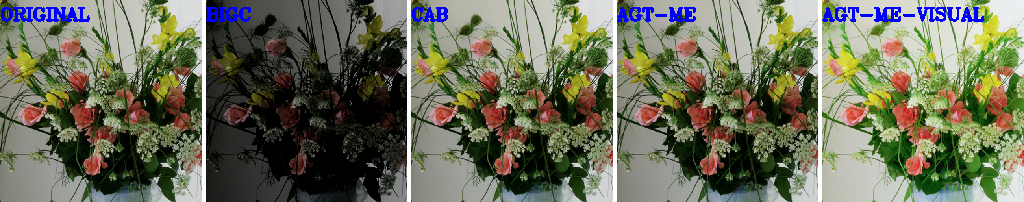}
}\hfil
\subfloat[Flower 2]{
\includegraphics[width=7.2in]{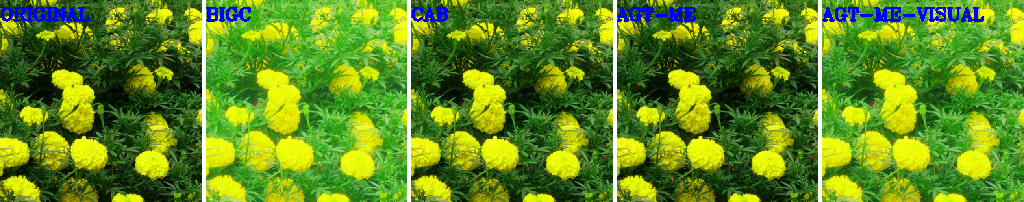}%
}\hfil
\subfloat[Boy]{
\includegraphics[width=7.2in]{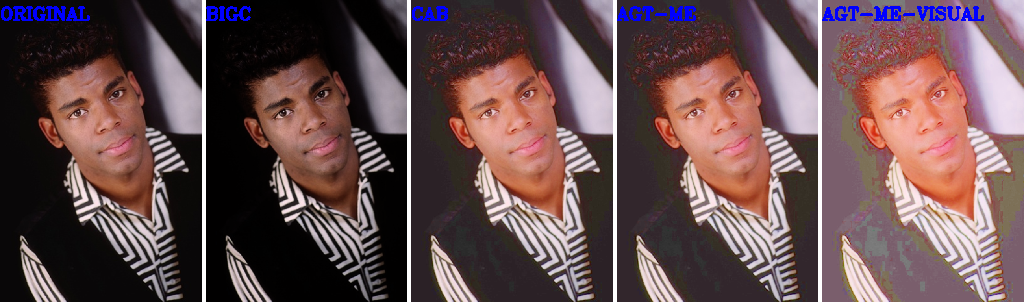}%
}\hfil
\subfloat[Lena]{
\includegraphics[width=7.2in]{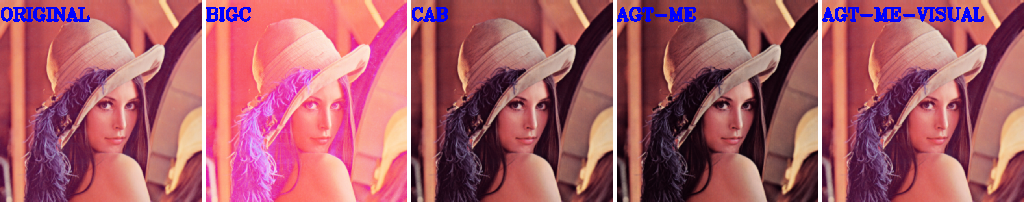}%
}\hfil
\subfloat[Airplane]{
\includegraphics[width=7.2in]{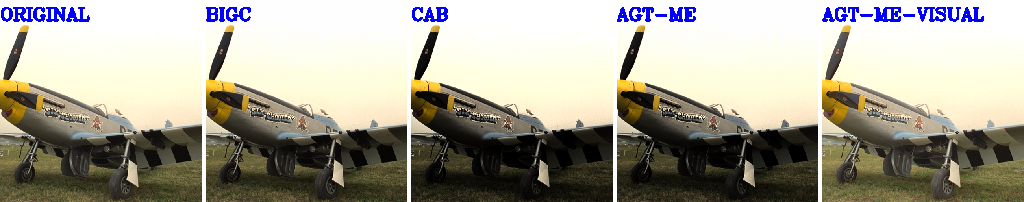}%
}
\caption{Image contrast enhancement results with different gamma correction methods on natural images.}
\label{fig_6}
\end{figure*}

Another important application of adaptive gamma transformation is image contrast enhancement (CE). The gamma transformation can be used alone for CE, or as a compositional part to form a complex CE algorithm along with other image processing algorithms\cite{arici2009histogram} (histogram equalization, log transformation, contrast stretch, etc). Because this paper focuses on adaptive gamma correction, only adaptive gamma transformation methods are compared in the application of natural and medical image contrast enhancement. Image contrast enhancement performance cannot be assessed with entropy metrics due to the \em{quantized entropy decrease barrier}\em{}. And assessing the visualized results by human experts is an important means to measure CE performance. As a result, we only provide visualized image results in this CE application.

The 127 natural images are adopted from CBSD68, Kodak24 \cite{franzen1999kodak,zhang2018ffdnet}, McMaster \cite{zhang2011color,zhang2018ffdnet}\footnote{https://github.com/cszn/FFDNet/tree/master/testsets}, a standard test set\footnote{http://www.imageprocessingplace.com}. Fig. \ref{fig_6} gives five images out of 127 test images and the corresponding results.
The BIGC method makes Flower 1 darker and Flower 2 / Lena brighter resulting in unnatural looks, which illustrates the instability. On the contrary, the CAB, AGT-ME, and AGT-ME-VISUAL demonstrate consistent performance. Both the CAB and AGT-ME methods give a natural-looking appearance. The AGT-ME-VISUAL increases the brightness of AGT-ME results. The hidden details or artifacts in the dark area become visible, resulting in more informative images with higher contrast. Meanwhile, the AGT-ME-VISUAL results suffer contrast over-enhancement more or less if the image has a very dark appearance (eg. Boy). Note that, the artifacts are always there instead of introduced by AGT-ME-VISUAL.

\begin{figure*}[!t]
\centering
\subfloat[]{\includegraphics[width=6.8in]{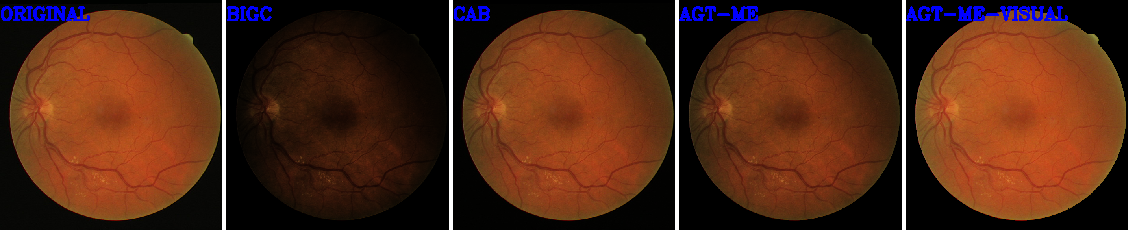}}
\hfill
\subfloat[]{\includegraphics[width=6.8in]{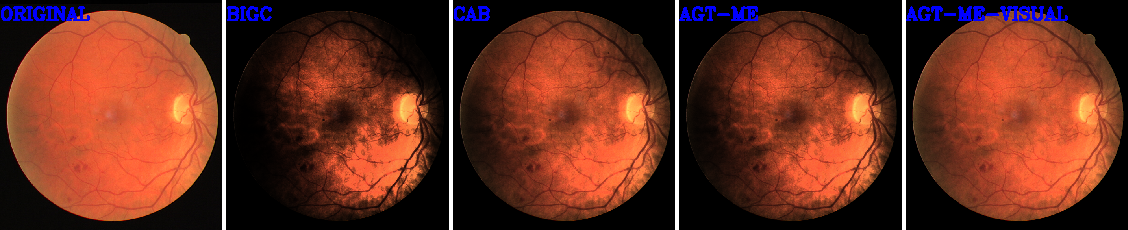}}
\hfill
\subfloat[]{\includegraphics[width=6.8in]{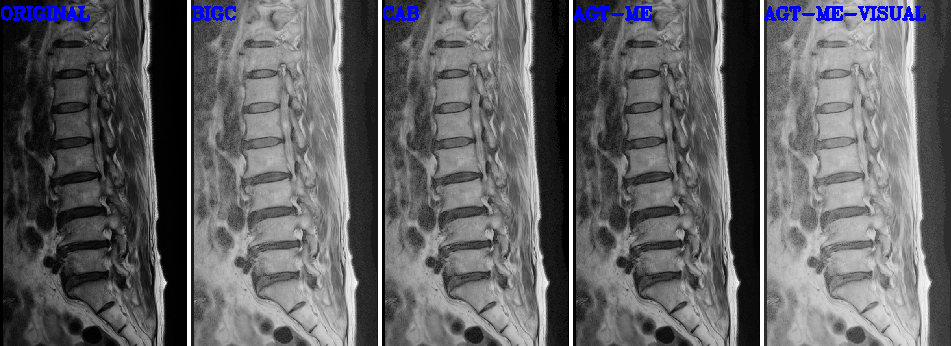}}
\hfill
\subfloat[]{\includegraphics[width=6.8in]{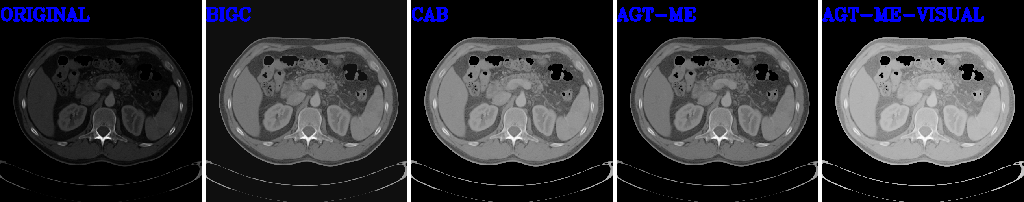}}
\hfill
\subfloat[]{\includegraphics[width=6.8in]{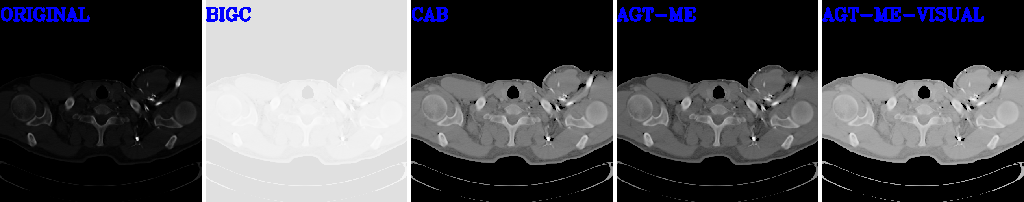}}
\caption{Image contrast enhancement results with different gamma correction methods on medical images.}
\label{fig_7}
\end{figure*}

Different from natural images, medical images have varied types, including ultrasound image, computed tomography scan image (CT), magnetic resonance image (MRI). Clear medical images are very important for the specialist diagnosis of related tissue lesions, such as diabetic retinopathy\cite{pratt2016convolutional}, tumor\cite{abul2008intradural} and cancer\cite{swensen2003lung}. In order to further investigate the contrast enhancement performance on medical images, we employed three relevant dataset - a retinal image dataset DRIVE \cite{niemeijer2004comparative,staal2004ridge}, a spinal MRI image dataset \cite{noauthor_mri_dataset_nodate}, and an abdominal CT image dataset \cite{noauthor_ct_nodate}. Fig. \ref{fig_7}(a) and (b) provide two raw color fundus images in DRIVE and corresponding CE results. Due to patients of varying ethnicity and age groups, the fundus images have extremely varied levels of lighting conditions. As a result, some diagnostic details in the original images may be concealed to some extent. Meanwhile, the significant white dots (Fig.\ref{fig_7}(a)) and dark spots (Fig.\ref{fig_7}(b)) are enhanced by AGT-ME-VISUAL, which is helpful to the funduscopic examination. Fig. \ref{fig_7}(c) represents a spinal MRI image\cite{noauthor_mri_dataset_nodate}. Fig. \ref{fig_7}(d) and (e) represent two abdomen CT images\cite{noauthor_ct_nodate}. The raw MRI/CT data is saved in a 64-bit float format, and it is not suitable for visual diagnosis without further enhancement. The AGT-ME-VISUAL provides satisfactory enhanced results by performing adaptive gamma transformation merely.

\subsubsection{Gamma Correction for FPP Image}

\begin{figure*}[!t]
\centering
\subfloat[]{\includegraphics[width=3.5in]{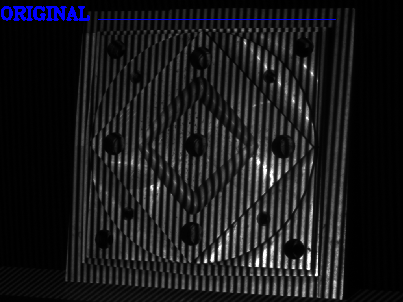}}
\hspace{0.01in}
\subfloat[]{\includegraphics[width=3.5in]{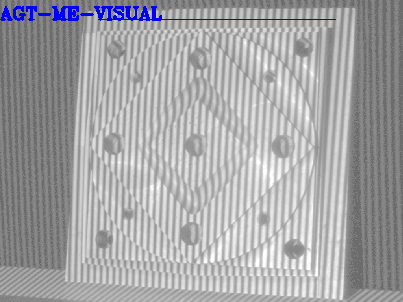}
}\hfil
\subfloat[]{\includegraphics[width=3.5in]{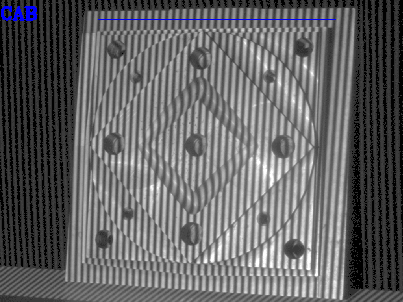}}
\hspace{0.01in}
\subfloat[]{\includegraphics[width=3.5in]{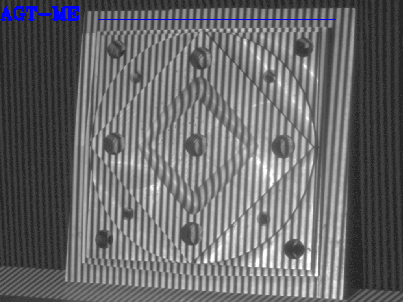}
}\hfil
\subfloat[]{\includegraphics[width=3.5in]{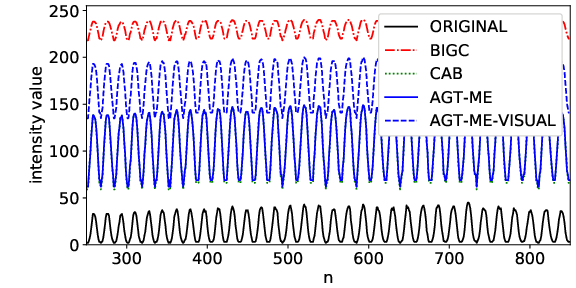}}
\hspace{0.01in}
\subfloat[]{\includegraphics[width=3.5in]{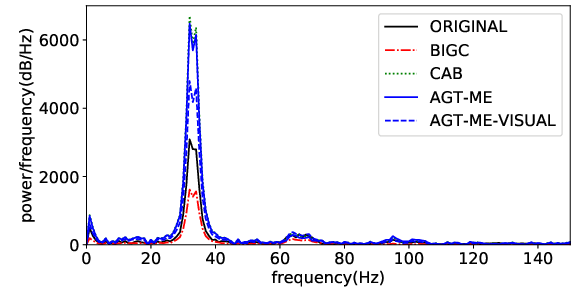}}
\caption{A FFP image and results. (a) an original FPP image; (b)-(d) results from different methods; (e) scan-lines from (a)-(d) at the blue lines; (f) power spectrum analysis of the scan-lines. Note the overlap of CAB and AGT-ME curves.}
\label{fig_8}
\end{figure*}


Gamma distortion is an unignorable factor that substantially affects the phase estimation, resulting in the FPP measurement accuracy reduction. Gamma correction is thus an essential image pre-processing procedure of FPP measurement. Here, an industrial workpiece FPP image\cite{xiong_design_2013} was utilized to evaluate the gamma correction on this FPP application. Fig. \ref{fig_8}(a)-(d) display an original image, AGT-ME-VISUAL image, CAB result, and AGT-ME result respectively. To attenuate the light reflection (intensity saturation), the original dark images were captured with a deliberate low exposure setting. The CAB, AGT-ME, and AGT-ME-VISUAL improved the image brightness. Further observing the intensity curves of scan-line (Fig.\ref{fig_8}(e)), The CAB and AGT-ME perform similarly and provide the largest amplitudes of the main harmonics. When a sinusoidal signal is distorted, new harmonics are introduced \cite{farid2001blind,liu2010gamma}.  Power spectrum analysis (Fig.\ref{fig_8}(f)) can clearly identify the unwanted harmonics near  $65Hz$ and $95Hz$ frequencies due to gamma distortion. The first-order harmonic ($\sim 33Hz$), the useful phase component in FPP, has been extremely amplified with CAB / AGT-ME method. It means that the quality of the FPP image has been significantly improved.

\section{Conclusion}
\label{conclusion}

An adaptive gamma transformation method, acronymed AGT-ME, is proposed to perform blind inverse gamma correction based on the wide-acknowledged image maximum entropy prior. The AGT-ME method successfully formulated the problem with maximizing differential entropy via change of variable rule, and further provided a mathematical concise closed-form solution of the optimization. It overcomes the \em{quantized entropy decrease barrier}\em{} and leads to an efficient algorithm implementation. Considering the human eye has a non-linear perception sensitivity, a modified version AGT-ME-VISUAL is also proposed. The characteristic insight of AGT-ME/ AGT-ME-VISUAL is gained with sufficient experimental evidence. The amount of gamma correction can be accurately identified with averaging RMSE $0.0439$. It costs no more than $100ms$ to deal with a $1024\times1024$ image. Three applications (automatic gamma adjustment, image contrast enhancement, fringe projection profilometry) demonstrate the wide potential application areas. Furthermore, a python implementation of the AGT-ME method is also shared and we recommend you use it when you need a fast and accurate gamma estimation.


%



\section*{Acknowledgment}
The authors thank the anonymous reviewers for their insightful comments and suggestions. This work was supported by 
National Natural Science Foundation of China (Grant Nos. 81900706 and 51705371), the
Natural Science Foundation of Jiangsu Province (Grant No. BK20180235).

\ifCLASSOPTIONcaptionsoff
  \newpage
\fi



\bibliographystyle{IEEEtran}
\bibliography{IEEEabrv,./BlindInverseGammaCorrectionwithMaximizedShannonInformationEntropy}

%



%

\begin{IEEEbiography}[{\includegraphics[width=1in,height=1.25in,clip,keepaspectratio]{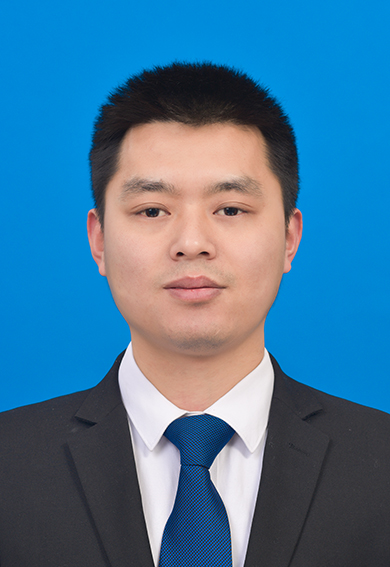}}]{Yong Lee}
received his BSc and Ph.D degrees from Huazhong University of Science and Technology (HUST), Wuhan, P. R. China, in 2012 and 2018 respectively. He worked at Cobot (Wuhan Cobot Technology Co., Ltd.) as an algorithm scientist from 2018 to 2019.  He is currently a Research Fellow with Department of Computer Science at National University of Singapore. His research interests include image processing, particle image velocimetry, machine learning and deep learning.
\end{IEEEbiography}


\begin{IEEEbiography}[{\includegraphics[width=1in,height=1.25in,clip,keepaspectratio]{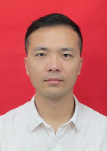}}]{Shaohua Zhang} received his BSc from Xi'an Jiaotong University, P. R. China, in 2007. He obtained his PhD degree from HUST, China, in 2016. He works as an algorithm scientist at Cobot(Wuhan Cobot Technology Co., Ltd.) since November 2016. His research interests include robotic vision and deep learning.
\end{IEEEbiography}

\begin{IEEEbiography}[{\includegraphics[width=1in,height=1.25in,clip,keepaspectratio]{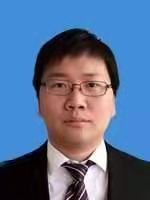}}]{Miao Li} received his BSc and MSc from HUST, China, in 2008 and 2011 respectively. He obtained his PhD in Robotics and Intelligent System at Ecole Polytechnique Federale de Lausanne (EPFL), Switzerland, in 2016. Since December 2016, he has been with Wuhan University, where he is currently an associate professor. His research interests include robotics, machine learning and applied nonlinear control, which encompass robot learning, robot object grasping and manipulation, human robot interaction, robotic hand and tactile sensing, and neuroscience. He is the author of about 30 papers in international/national journals and conferences. His PhD work was awarded for EPFL PhD Thesis Award in 2018, with his work on the closed-loop grasping and object manipulation. He is the finalist of T.J. Tarn Best Paper Award at ROBIO 2016 and he received the second place at IROS 2016 Robotic Grasping and Manipulation Challenge. Miao Li is also a co-founder of a start-up working on intelligent industry robots for real-world problems.
\end{IEEEbiography}


\begin{IEEEbiographynophoto}{Xiaoyu He}received her BS. degree in 2012 from Hubei Minzu University, P.R. China, and obtained her Ph.D degree in Endocrinology at Tongji Medical College, HUST, Wuhan, P. R. China in 2017. Since 2017, she is a clinical physician in the Department of Endocrinology, Tongji Hospital, Tongji Medical College, HUST, Wuhan, P. R. China. Her research interests include medical data analysis, diabetes and SUMOylation.

\end{IEEEbiographynophoto}





\end{document}